# Three-dimensional Atmospheric Dynamics of Jupiter from Ground-based Doppler Imaging Spectroscopy in the Visible

Francois-Xavier Schmider[1], Patrick Gaulme[2,3,4], Raúl Morales-Juberías[4], Jason Jackiewicz[3], Ivan Gonçalves[1,5], Tristan Guillot[6], Amy A. Simon[7], Michael H. Wong[8], Thomas Underwood[9], David Voelz[9], Cristo Sanchez[3], Riley DeColibus[3], Sarah A. Kovac[3,10], Sean Sellers[3], Doug Gilliam[11], Patrick Boumier[12], Thierry Appourchaux[12], Julien Dejonghe[1], Jean Pierre Rivet[1], Steve Markham[1,3], Saburo Howard[1,13], Lyu Abe[1], Djamel Mekarnia[1], Masahiro Ikoma[14], Hidekazu Hanayama[14], Bun'ei Sato[15], Masanobu Kunitomo[1,16], and Hideyuki Izumiura[17]

[1] Université Côte d'Azur, Observatoire de la Côte d'Azur, Laboratoire Lagrange, UMR7293 CNRS, 06304 Nice, France; schmider@oca.edu
[2] Thüringer Landessternwarte, Sternwarte 5, 07778 Tautenburg, Germany
[3] Department of Astronomy, New Mexico State University, P.O. Box 30001, MSC 4500, Las Cruces, NM 88003-8001, USA
[4] New Mexico Tech, Department of Physics, 801 Leroy Place, Socorro, NM 87801, USA
[5] Université Paul Sabatier, Institut de Recherche en Astrophysique et Planetologie, Observatoire Midi-Pyrénées, Toulouse, France
[6] Université Côte d'Azur, Observatoire de la Côte d'Azur, Lagrange, Nice, France
[7] NASA Goddard Space Flight Center, Solar System Exploration Division, 8800 Greenbelt Road, Greenbelt, MD 20771 USA
[8] University of California, Center for Integrative Planetary Science, 501 Campbell Hall, Berkeley, CA 94720-3411, USA
[9] Klipsch School of Electrical and Computer Engineering, New Mexico State University, MSC 3-O, Goddard Annex 160B, Las Cruces, NM 88003, USA
[10] Southwest Research Institute, 1050 Walnut Street, Suite 300, Boulder, CO 80302, USA
[11] Sunspot Solar Observatory, P.O. Box 62, Sunspot, NM 88349, USA
[12] Université Paris-Saclay, Institut d'Astrophysique Spatiale, UMR 8617, CNRS, Bâtiment 121, 91405 Orsay Cedex, France
[13] Institute for Computational Science, University of Zurich, Winterthurerstr. 190, CH8057 Zurich, Switzerland
[14] Division of Science, National Astronomical Observatory of Japan (NAOJ), Mitaka, Tokyo 181-8588, Japan
[15] Department of Earth and Planetary Sciences, School of Science, Institute of Technology, 2-12-1 Ookayama, Meguro-ku, Tokyo 152-8551, Japan
[16] Department of Physics, Kurume University, 67 Asahimachi, Kurume, Fukuoka 830-0011, Japan
[17] NAOJ, Okayama Observatory, Okayama, Japan

Received 2023 December 12; revised 2024 February 27; accepted 2024 February 28; published 2024 April 16

## Abstract

We present three-dimensional (3D) maps of Jupiter's atmospheric circulation at cloud-top level from Doppler-imaging data obtained in the visible domain with JIVE, the second node of the JOVIAL network, which is mounted on the Dunn Solar Telescope at Sunspot, New Mexico. We report on 12 nights of observations between 2018 May 4 and May 30, representing a total of about 80 hr. First, the average zonal wind profile derived from our data is compatible with that derived from cloud-tracking measurements performed on Hubble Space Telescope images obtained in 2018 April from the Outer Planet Atmospheres Legacy program. Second, we present the first ever 2D maps of Jupiter's atmospheric circulation from Doppler measurements. The zonal velocity map highlights well-known atmospheric features, such as the equatorial hot spots and the Great Red Spot (GRS). In addition to zonal winds, we derive meridional and vertical velocity fields from the Doppler data. The motions attributed to vertical flows are mainly located at the boundary between the equatorial belts and tropical zones, which could indicate active motion in theses regions. Qualitatively, these results compare well to recent Juno data that have unveiled the 3D structure of Jupiter's wind field. To the contrary, the motions attributed to meridional circulation are very different from what is obtained by cloud tracking, except at the GRS. Because of limitations with data resolution and processing techniques, we acknowledge that our measurements of the vertical or meridional flows of Jupiter are still to be confirmed.

*Unified Astronomy Thesaurus concepts:* Jupiter (873); Planetary atmospheres (1244); Doppler imaging (400)

*Supporting material:* machine-readable table

## 1. Introduction

The atmospheric dynamics of the giant planets, in particular Jupiter, are characterized by the presence of strong, alternating zonal wind jets. Historically, the zonal wind profile of Jupiter has been measured by tracking cloud structures on images of the planet separated in time by at least one rotation period (Limaye 1986; García-Melendo & Sánchez-Lavega 2001; Porco et al. 2003). Multiple studies have used this cloud-tracking technique to characterize the evolution of the zonal wind profile as a function of time. Changes were observed and have been associated with specific dynamical features such as the Great Red Spot (GRS) and other vortices or atmospheric features (Read et al. 2006; Barrado-Izagirre et al. 2013; Hueso et al. 2017; Johnson et al. 2018). In addition, high-resolution images acquired by the Hubble Space Telescope (HST) and space missions such as Voyager or Cassini have allowed us to characterize the meridional component of the wind field from two-dimensional (2D) correlations of these images. These results have unveiled the role of eddies in pumping energy into the zonal jets and provided the energy spectrum of the 2D flow (Salyk et al. 2006; Choi & Showman 2011; Galperin et al. 2014; Tollefson et al. 2017; Ingersoll et al. 2021). However, very little is still known about the detailed three-dimensional

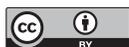

(3D) structure of the wind field of the atmospheres of these planets and how it is connected to the observed 2D wind field.

In 1995, the Doppler Wind Experiment and accelerometers on board the Galileo descent probe provided in situ measurements of the variations of these zonal winds as a function of depth. Both sets of measurements revealed wind speeds at the cloud tops (≈700 mbar level) that were in agreement with the results of cloud tracking (80–100 m s$^{-1}$) at the probe entry site (6°.5 north). Below the cloud level, the winds increased dramatically up to ≈170 ms$^{-1}$ at ≈4 bars. Below this level, the winds remained nearly uniform down to the 21 bar level, where the probe stopped emitting signals (Atkinson et al. 1997; Seiff et al. 1997). More recently, detailed analyses of the ammonia abundance and gravity field measurements, obtained by the Juno spacecraft orbiting Jupiter in close polar orbits, have been used to characterize the 3D structure of the zonal winds (Guillot et al. 2018; Kaspi et al. 2018; Duer et al. 2021; Fletcher et al. 2021) and the vortices embedded in them (Bolton et al. 2021; Parisi et al. 2021). The picture that emerges from these studies is that Jupiter's banded appearance is caused by upwelling and downwelling cells similar to the Ferrel cells on Earth. Juno measurements, when combined with a simple advection-relaxation model, allow for characterizing the spatial structure of the zonally averaged velocity field associated with the cells (Duer et al. 2021). However, it is impossible to deduce the absolute values of the velocities associated with the meridional and vertical transport in the cells.

Doppler velocimetry has long been considered both for the search of planetary oscillations and for measuring atmospheric dynamics (Vorontsov et al. 1976; Schmider et al. 1991; Gaulme et al. 2011). The best approach to track the atmospheric motions in the visible domain—vertical for seismic observations, horizontal for wind circulation—consists of measuring the Doppler shift of solar Fraunhofer lines that are reflected by the planet's upper cloud layers, as the Doppler signal is enhanced by reflection (Gaulme et al. 2018). Regarding the seismology of giant planets, all the attempts have been dedicated to Jupiter because it is the biggest and brightest target seen from Earth. The first observations with a magneto-optical filter (MOF; Cacciani & Fofi 1978) were led by Schmider et al. (1991), then followed by observations with a Fourier transform spectrometer (Mosser et al. 1993, 2000), a double MOF (Cacciani et al. 2001), and the first dedicated instrument, SYMPA (Schmider et al. 2007; Gaulme et al. 2008, 2011), also a Fourier transform spectrometer. Observations by different groups (Schmider et al. 1991; Mosser et al. 1993, 2000; Gaulme et al. 2011) concluded the presence of oscillations at a low signal-to-noise level, with amplitude between 0.1 and 1 ms$^{-1}$. Regarding atmospheric dynamics, most of the efforts have been dedicated to Venus, in particular, to support the ESA Venus Express mission (Lellouch et al. 2008). Venus observations were mostly performed by scanning the planet with a single fiber-fed high-resolution spectrograph (Widemann et al. 2008; Machado et al. 2017, and references therein) or with long-slit spectrographs (Machado et al. 2012; Gaulme et al. 2019).

The first zonal wind profile of Jupiter measured with Doppler velocimetry was obtained with the prototype of the Doppler Spectro Imager employed in the present work, which is an imaging spectrometer inherited from SYMPA (Gonçalves et al. 2016; Soulat et al. 2017). The observations were led in 2016 at the Calern observatory in southern France (Gonçalves et al. 2019). The zonal profile derived from that data set revealed significant discrepancies with the cloud-tracking profiles in two specific regions, namely, the North Equatorial Belt and the northern part of the Equatorial Zone. Recently, another instrument was used to derive velocities on Jupiter from Doppler images based on a potassium MOF (Shaw et al. 2022). Observations obtained during 6 weeks in Hawaii with a 3.6 m telescope in very good (≈0″.85) seeing conditions were used to derive a zonal wind profile. Although the measurements had to be filtered to remove a low spatial frequency bias, they derived a zonal wind profile that exhibits many small-scale details comparable to previously published work based on cloud-tracking data (Galperin et al. 2001). Finally, Machado et al. (2023) reported zonal wind measurements of Jupiter conducted with the ESPRESSO high-resolution spectrograph at the Very Large Telescope observatory (Pepe et al. 2021). This was an exploratory effort aimed at investigating the effectiveness of measuring winds on Jupiter using high-resolution spectroscopic data obtained with ground-based telescopes. Within the limited spatial (±20° in latitude from the equator) and temporal (2 nights) coverage, their zonal wind results, albeit tentative, are mostly consistent with previous measurements, thus validating the effectiveness of the technique.

In this paper, we report the first 3D map of Jupiter's atmospheric circulation ever obtained with imaging spectroscopy in the visible. The JOVIAL network was set up between 2016 and 2019 with three Doppler Spectro Imagers placed on three telescopes around the world, in France, Japan, and the USA (Schmider et al. 2013). The data that we use in this paper were obtained in 2018 with the second node of the JOVIAL network—the Jupiter Interferometric Velocity Experiment in New Mexico (JIVE in NM, hereafter JIVE)—which is mounted at the focus of the Dunn Solar Telescope (DST) in Sunspot, New Mexico (Underwood et al. 2017). Thanks to an improved data analysis pipeline with respect to the work reported by Gonçalves et al. (2019), we were able to produce a full 3D wind field of the planet on top of the mean zonal wind profile. Our new results point out detectable vertical motions at the latitudes that separate the equatorial belts and the tropical zones.

## 2. Observations

### 2.1. JIVE Velocimetry Data

The velocimetry data were obtained with the JIVE instrument on the DST from 2018 May 4 to May 31. The DST consists of a turret located at the tip of a 40 m high tower with an entrance window and a flat mirror that sends the beam to the main mirror located 60 m below ground level (Zirker 1998). The focal ratio is 72, and the effective aperture is 76 cm, despite the primary mirror's diameter being 1.63 m. Several ports are available for observations, each corresponding to a different instrument.

The JIVE instrument was delivered by Observatoire de la Côte d'Azur to New Mexico State University at the end of 2017 and was then installed on an optical bench inside the main room of the telescope. We refer the reader to Underwood et al. (2017) for details about JIVE's installation at Sunspot. In 2018 May, we were granted an entire month of nighttime observations for the opposition of Jupiter. Ultimately, we were able to observe for 12 nights, representing a total of 66.5 hr of

**Table 1**
Sunspot Observation Summary in 2018 May

| Date | Duration (hr) | N Points | Diameter (arcsec) | Phase (deg) | Mean Seeing (arcsec) | Flux (1e$^8$ photons) | Noise (m s$^{-1}$) |
|---|---|---|---|---|---|---|---|
| 2018-05-04 03:29:30 | 06:12:29 | 693 | 44.7480 | 0.9918 | 2.5374 | 7.2851 | 3.9403 |
| 2018-05-05 02:51:30 | 08:35:29 | 1007 | 44.7658 | 0.7926 | 2.4283 | 7.0466 | 4.0065 |
| 2018-05-06 03:23:00 | 07:19:29 | 788 | 44.7801 | 0.6032 | 2.1001 | 7.3475 | 3.9236 |
| 2018-05-07 05:50:00 | 04:22:00 | 331 | 44.7922 | 0.4117 | 1.7709 | 5.1279 | 4.6966 |
| 2018-05-09 03:01:30 | 07:45:00 | 863 | 44.8051 | 0.2440 | 1.7890 | 7.1807 | 3.9689 |
| 2018-05-10 02:55:30 | 07:32:30 | 865 | 44.8073 | 0.3517 | 1.5720 | 6.7095 | 4.1059 |
| 2018-05-15 04:16:00 | 05:46:29 | 415 | 44.7715 | 1.3084 | 1.6908 | 7.4500 | 3.8965 |
| 2018-05-16 03:05:30 | 04:08:30 | 480 | 44.7571 | 1.4866 | 1.7556 | 7.4679 | 3.8918 |
| 2018-05-17 02:37:30 | 06:49:30 | 789 | 44.7373 | 1.6934 | 1.6803 | 7.3554 | 3.9215 |
| 2018-05-24 02:35:30 | 07:17:00 | 713 | 44.5196 | 3.0798 | 2.5095 | 6.6184 | 4.1340 |
| 2018-05-25 02:33:00 | 07:12:30 | 797 | 44.4784 | 3.2677 | 1.7993 | 7.0195 | 4.0142 |
| 2018-05-31 02:59:00 | 04:07:30 | 343 | 44.1699 | 4.3927 | 1.8789 | 7.4520 | 3.8960 |

**Note.** The table provides the date, duration, and quality of the observations and a summary of Jupiter's observability (apparent diameter and phase). Flux is the mean number of photons received in 30 s. Noise is the standard deviation of the mean velocity for each image. The values are close to the theoretical photon noise level.

observations. Table 1 displays the different nights with the conditions of the observations.

A major improvement of the present measurements with respect to those led at the Calern observatory (Gonçalves et al. 2019) comes from the fact that the DST rotates around its azimuth axis, allowing us to take images with multiple orientations of the instrument with respect to Jupiter (Underwood et al. 2017). That way, we could calibrate instrumental biases related to the orientation of the images on the detector that were identified from Calern data. Another improvement with respect to Gonçalves et al. (2019) is the seeing quality, which was ≈1″.9 on average, and generally between 1″.3 and 2″, whereas it was between 2″ and 3″.5 during observations in Calern.

### 2.2. HST Outer Planet Atmospheres Legacy Data

To interpret our results and support the processing of the JOVIAL data, we use HST data obtained by the Outer Planet Atmospheres Legacy (OPAL) program (Simon et al. 2015). In particular, we used Jupiter images acquired with WFC3/UVIS on 2018 April 17 (Cycle 25). The images were processed using an ellipsoid limb-fitting technique with equatorial and polar radii of 71,492 and 66,854 km, respectively. A Minnaert correction (ratio of the cosines of the incidence and emission angles) to the power of a limb-darkening coefficient was applied to remove limb darkening and produce cylindrical coordinate maps. Each map is generated at the sub-Earth longitude ±39.°9 at 10 pixels deg$^{-1}$ resolution, between −79.°8 and +79.°8 planetographic latitude, and can be mosaicked to cover 360° of longitude, with seams between maps interpolated to smooth where needed. Two global maps exist for most filters, covering two full rotations of Jupiter.[18]

## 3. Data Processing

### 3.1. Cloud Tracking from HST Data

To derive the zonal wind profiles from the HST observations, we made use of the individual pairs of maps rather than the global maps. The advantage of doing so is that the exact time between images is known, which is crucial for accurately estimating the wind speed. Consecutive maps in a given filter were roughly spaced by one Jovian rotation but still overlapped in longitude, typically from ≈25° to ≈70°.

We derived the zonal velocities by using a one-dimensional cross-correlation technique as described in previous studies (e.g., Johnson et al. 2018). This technique consists of scanning the maps in longitude for each latitude and cross-correlating the signals obtained this way after removing the mean and dividing by the standard deviation. Once the displacement obtained from the cross-correlation is obtained, we divide this displacement by the time difference between the images to estimate the zonal wind speed at each latitude. Not all the image pairs produce clean zonal profiles for all latitudes using this technique. For example, image pairs overlapping less than 30° usually produced profiles that were unrealistically high or low. These wind profiles were not considered for averaging the final zonal wind profile.

To reduce the noise, the profiles derived from individual image pairs were filtered before averaging them into the final zonal wind profile. For this, we employed a Savitzky–Golay filter, which is designed for smoothing spectral line data without degrading the lines' height or width (Savitzky & Golay 1964). The standard deviation of the series of individual zonal profiles around the average profile is what is considered to be the error shown in the figures.

### 3.2. Extracting Velocities from JIVE Data

#### 3.2.1. Disentangling Velocities from Photometry and Point-spread Function

Doppler velocimetry observations are difficult to reduce and interpret. JIVE is designed to produce simultaneously images of the planet and maps of the average Doppler shift of a set of solar absorption lines in the visible domain that are reflected by the uppermost cloud layers of Jupiter ($P \approx 500$–700 hPa) (Figure 1). Such maps are commonly referred to as "dopplergrams." Contrary to the abovementioned Doppler measurements obtained with single fiber-fed or long-slit spectrographs, JIVE produces a complete dopplergram of the planet at each exposure. From the projected velocity at a given point of the planetary surface, it is possible to infer both the motion of the surface and its variations as a function of time.

---
[18] https://archive.stsci.edu/prepds/opal/

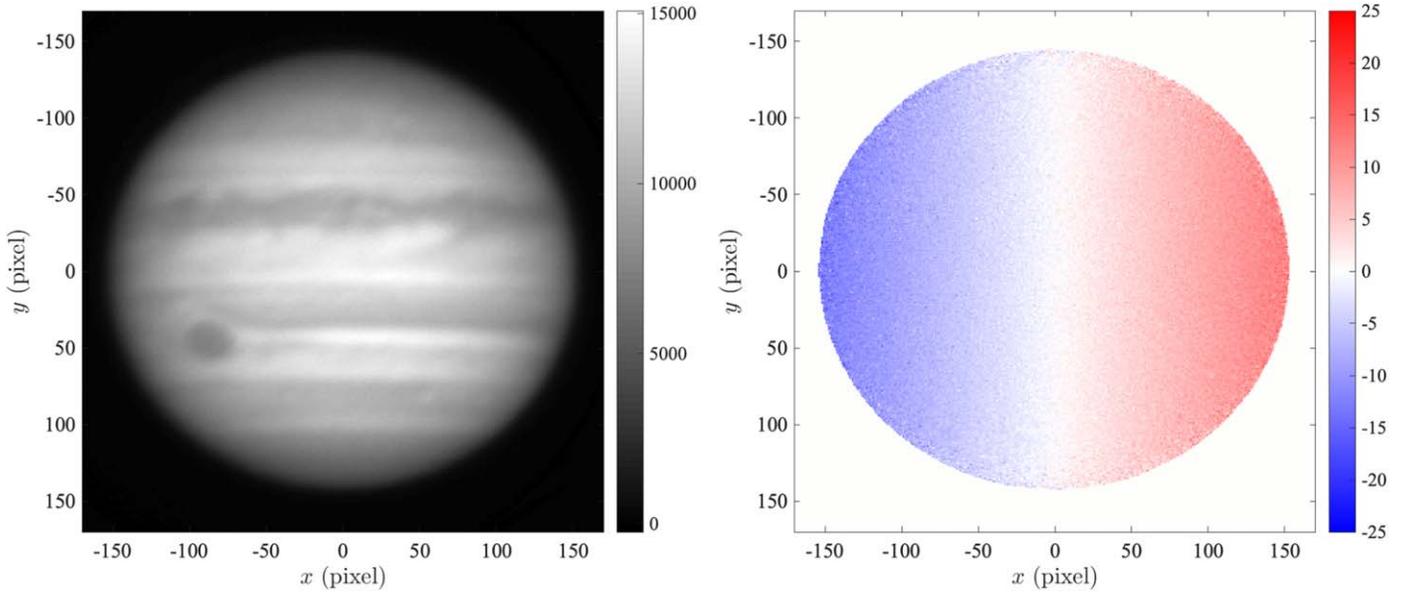

**Figure 1.** Jupiter observed by JIVE on 2018 May 17 at the DST. Left panel: image of Jupiter that corresponds with an exposure time of 30 s, where the flux is expressed in photons pixel$^{-1}$. Right panel: line-of-sight velocity map corresponding to the same acquisition, aka dopplergram, where the velocity is expressed in km s$^{-1}$. The $x$- and $y$-axes are expressed in pixels, and the seeing was estimated to be about 1″4.

The general calibration and data processing are detailed in Gonçalves et al. (2019).

At a given point on the planet, the measured Doppler shift integrates the velocity projected toward the source (Sun) and toward the observer (Earth). The sum of these two shifts gives the projection factors of the individual wind components at that point. However, as originally pointed out by Civeit et al. (2005), the point-spread function (PSF), which includes the response of the focused optical imaging system and the atmospheric seeing, alters the radial velocity measurements because it blends regions with nonuniform Doppler shift and nonuniform photometry. The measured line-of-sight velocity is the convolution of the line-of-sight velocity with the photometric map of the considered object, including its degradation by the PSF. The measured Doppler signal $v_m$ measured in a given point can be expressed as

$$v_m = \frac{(F\, v_d) * P}{F * P}, \quad (1)$$

where $F$ is the local photometric flux on the planet, $v_d$ is the Doppler velocity, $P$ is the PSF, and the asterisks indicate the convolution. Therefore, a simple extraction of the Doppler signal from the JIVE data is not enough. A necessary step is to estimate the terms $F$ and $P$ of Equation (1) to extract the best possible estimator of $v_d$ out of $v_m$.

To better understand what is expected, we simulated the difference between $v_m$ and $v_d$ (Figure 2). For this, we built a photometric map of Jupiter $F$ from a 2018 May HST OPAL planisphere. Regarding the Doppler signal $v_d$, we assumed a simple solid-body rotator. We then computed the degraded dopplergram by assuming a Gaussian PSF $P$ with a FWHM of 2″. The difference $|v_m - v_d|$ is maximum toward the edge of the planet, where it reaches about 500 m s$^{-1}$, because both the photometric flux and the projected velocity vary rapidly. We see how important it is to correct our raw dopplergrams for this effect. In particular, two results deserve to be remembered. First, we see that fake Doppler shifts are generated by a sharp flux variation, such as that surrounding the GRS. Second, the zone-and-belt alternating structure that is clearly visible on the photometric map does not bias the dopplergram, despite sharp contrast variations. This is because the main contribution to the blend between photometry and velocity field is the coupling between the solid-body rotation and the photometry ($\pm 12.5$ km s$^{-1}$ along the equator versus a few m s$^{-1}$ expected for the meridional or vertical flows). Significant banded structure in the bias map displayed in Figure 2 would have appeared in the case of km s$^{-1}$ flows in the meridional or vertical directions.

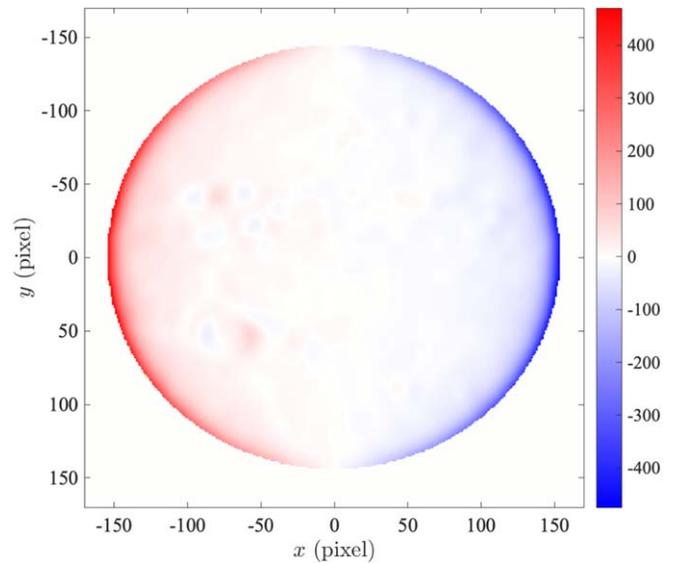

**Figure 2.** Simulation of the difference of velocity between a solid rotation model and its corresponding Doppler map that includes the degradation of the velocity signal by the PSF. The PSF is assumed to be Gaussian with an FWHM of 2″. The flux model employed to simulate the data is based on an actual high-resolution image of Jupiter. The color scale indicates the velocity in m s$^{-1}$.

Unfortunately, a direct deconvolution of the dopplergrams is not achievable in a simple manner. Indeed, recovering the velocity map from the measured Doppler map would entail a high-resolution photometric map $F$, which is actually not measured. Even a diffraction-limited image of Jupiter from HST would not work, for two reasons. The first is that Jupiter's atmosphere is in constant evolution, and an image taken several weeks or even days earlier would not give the appropriate reference. Second, the HST imaging system has filters that do not correspond to JIVE's entrance filter (519.4 nm), and the photometric features of Jupiter's upper atmosphere strongly depend on the optical wavelength (e.g., Dahl et al. 2021). We considered having a separate lucky-imaging device next to the DST to get simultaneous diffraction-limited images of Jupiter, but it would not work as well, because lucky imaging requires much flux, and hence broad optical filters, for getting exposures shorter than a tenth of a second. Therefore, the photometric reference obtained by lucky imaging would not be appropriate for JIVE because of the differences in photometry between different bandpasses.

Fortunately, a compromise is possible that reduces most of the alteration of the velocity measurements by the PSF. In Appendix A, we show that the effect of the PSF on the velocity field can be reduced by using the blurred images that JIVE produces, coupled with a model of the PSF and a model of the zonal rotation velocity based on cloud tracking. Briefly, the approach consists of approximating the flux $F$ and Doppler velocity $v_d$ to first order at a given point on the image and replacing them in Equation (1). From Equation (A16), the estimated line-of-sight velocity $\widehat{v_d}$ can be extracted from the measured map $v_m$ at a point of coordinates $(x, y)$:

$$\widehat{v_d} \approx v_m - \xi \frac{\sigma_P^2}{F_m}\left(\frac{\partial F_m}{\partial x}\frac{\partial v_d}{\partial x} + \frac{\partial F_m}{\partial y}\frac{\partial v_d}{\partial y}\right), \quad (2)$$

with the help of the measured flux $F_m = F * P$, an estimate of the standard deviation $\sigma_p$ of the Gaussian PSF, and an ad hoc factor $\xi = 1.5$ arising from the numerical simulations performed in Appendix A.

As indicated in Appendix A, $v_d$ is a model of the Doppler velocity including the fast solid rotation and a zonal wind profile extracted from the HST/OPAL data. Regarding the photometry, we acknowledge that employing $\partial F_m/\partial x$ and $\partial F_m/\partial y$ instead of $\partial F/\partial x$ and $\partial F/\partial y$ is an approximation that limits the efficiency of our correction but it still improves the data quality.

Regarding the PSF, we assume it to be a Gaussian function, whose standard deviation $\sigma_P$ is estimated from JIVE's images. As done by Gonçalves et al. (2019), we estimate $\sigma_P$ by calculating the size of the image of Jupiter above a given threshold and comparing it with the theoretical size known from the ephemeris and the characteristics of the optical configuration. We note that the scale of the image on the sky could not be known to be better than 1%, so this method provides an estimate of the PSF size which could be biased. In addition, the PSF could be anisotropic because of optical aberrations in the telescope and in the instrument, which would contribute to altering the results.

From the corrected dopplergrams, it is possible to extract the zonal wind profile and the 2D zonal map with high confidence and the meridional and vertical 2D maps with lower confidence. Indeed, meridional and vertical projection factors, unlike the zonal ones, are mainly symmetrical around the central longitude. As a result, contamination of the Doppler signal by photometry will not be averaged out by the rotation of the planet.

### 3.3. From Dopplergrams to the Zonal Wind Profile

In what follows, we will only look to data between 68°S and 61°N, as measurements at higher latitude are too noisy to be considered. The inclination of Jupiter explains the asymmetry of the selected range. Before computing deprojected 2D velocity maps, we extract the mean zonal wind profile that we compare with that deduced from cloud tracking. For this, we follow the approach developed by Gonçalves et al. (2019), which consists of fitting the dopplergrams latitude by latitude. The dopplergram of a solid-body rotator observed at opposition (Sun–Earth–Jupiter aligned) from the equatorial plane of Jupiter would simply be a plane tilted from west to east (e.g., Gaulme et al. 2018). Jupiter is not exactly observed in such conditions: we are off the equatorial plane by about 3°, and the phase angle ran from 0°.2 to 4°.4 during the observing campaign. Since the departure to such conditions is small, we employ the method employed by Gonçalves et al. (2019), which consists of redressing the dopplergrams by interpolating them on a map where we have one latitude per row on the image. Since it is not a solid-body rotator, we fit a linear polynomial line by line, instead of a plane on the whole map. The mean slope of the Doppler velocities gives the zonal wind value. Dopplergrams of 3 minute sequences are averaged, and the individual zonal profiles are eventually averaged over the whole data set. Regarding error bars, we estimate the standard deviation from the individual fittings. Then, the resulting standard deviation contains both the noise level for the measurement and possible variations in the zonal wind.

Even though the data quality was improved with respect to Gonçalves et al. (2019), there are imperfections caused by instrumental artifacts that appear in the form of distortions of the line-of-sight velocity map. As in Gonçalves et al. (2019), we treat these residual distortions by subtracting a 2D second-degree polynomial fit from each individual dopplergram. This implies that the mean of the zonal profile is zeroed by our processing. Actually, the mean zonal wind value derived from cloud tracking is not 0 because it assumes a rotation rate as a reference. For the giant planets, the typical reference frame is the magnetic field rotation rate, which is presumably tied to the interior rotation. Therefore, to be consistent with cloud-tracking results, we offset the zonal profile such that both profiles coincide on average at high latitudes, where no significant differential rotation is measured.

Once the profile is offset, it is still impossible to directly compare our results with cloud tracking. Indeed, the cloud-tracking profile from HST/OPAL data has a higher spatial resolution (more pixels) and is almost diffraction-limited (no blurring). To quantify the impact of the PSF on the zonal wind profile, we simulate dopplergrams based on actual HST data that we degrade to match the observing conditions of JIVE. In practice, it involves considering the photometry and the velocimetry separately. Regarding photometry, we compute $F_m$ by convolving the HST image with a Gaussian function with a standard deviation equal to that estimated from JIVE images. Regarding velocimetry, we build a velocity map of the same size as the HST image by assuming the reference rotation rate plus the differential rotation, i.e., the zonal wind profile. We then degrade the simulated velocity map by applying Equation (1). That profile is directly comparable to that obtained with JIVE.

Comparing our Doppler profile to a degraded zonal profile is frustrating because we aim at deriving the "true" zonal wind velocity from our data. To get a good idea of what we would get in the absence of seeing alteration, we can simply add the difference between the original HST profile and the degraded HST profile. On the one hand, we acknowledge that we introduce information that does not belong to our data but to the HST profile. It artificially increases the spatial resolution of our results. On the other hand, the difference between the original profile and the degraded one does not depend much on the initial profile if it would have the same resolution as the final measurement. So, the process is not too wrong and permits us to see what values come out of our measurements. Alternatively, we could increase the resolution with some sort of deconvolution but with the effect of increasing the noise; we chose not to follow this path.

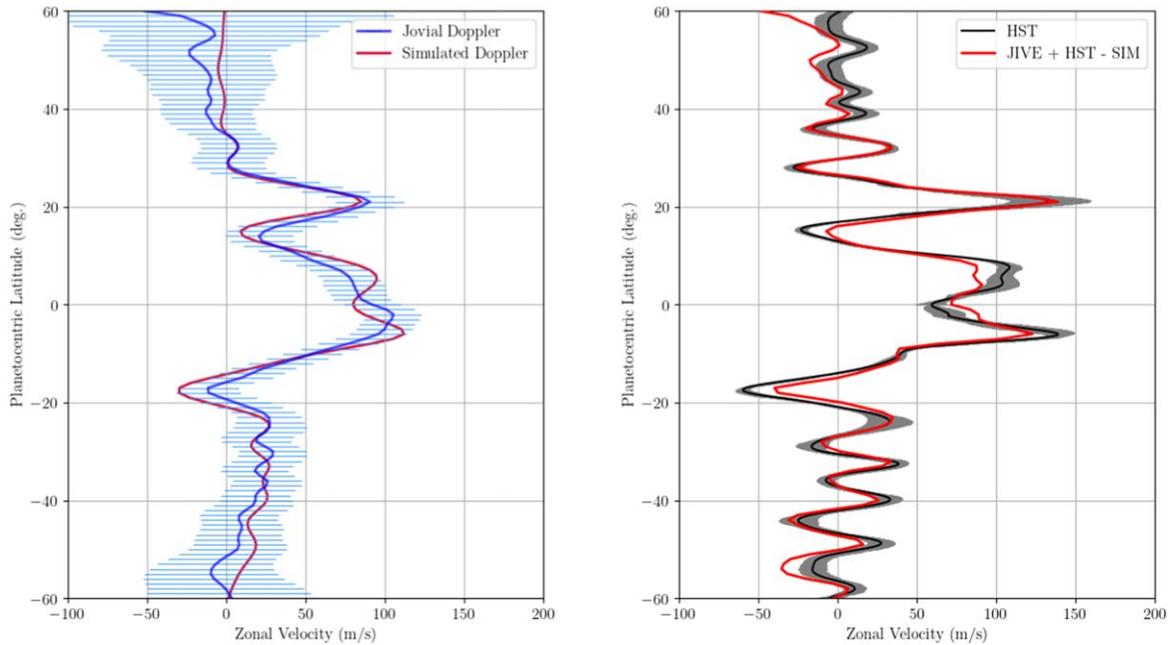

**Figure 3.** Left panel: mean zonal wind profile from Doppler measurements (in blue) compared to the simulated Doppler wind profile (in crimson). As in Gonçalves et al. (2019), the width of the dark blue line represents the theoretical noise level estimated from the known photon noise and the sensitivity, while the light blue bars represent the dispersion of all the measurements, which could include actual velocity variations, so it is an upper limit of the true noise level. Right panel: comparison between the wind profile obtained with the cloud-tracking technique on the 2018 HST OPAL data (black line) and the JIVE profile to which we added the difference between the HST profile and the simulated profile based on the HST profile (red line) The light gray area around the black line represents the error on the HST profile.

### 3.4. From Dopplergrams to 2D Wind Maps

Creating maps of atmospheric circulation entails deprojecting—i.e., dividing—the dopplergrams by the projection factors that transform velocity fields on the planet into line-of-sight velocity fields on the sky plane. That being said, such an operation cannot be directly performed without carefully taking the measurement noise into account (e.g., Gaulme et al. 2018).

For example, the dopplergrams are insensitive to zonal velocities along the bisector meridian of the planet, located halfway between the longitudes that point at the Sun and the observer, because the motion is perpendicular to the line of sight at that location. This is particularly visible in Figure 1, where the raw dopplergram is dominated by Jupiter's rotation. Similarly, the meridional component is affected by a high noise level near the equator since the observations are obtained from the Earth, which almost lies in the equatorial plane of Jupiter. At last, the vertical motion can be recovered around the equator but becomes noisier toward high latitudes.

To deproject the dopplergrams and assemble them into planispheres, it is then necessary to weight the deprojected maps when averaging them:

$$v_c = \frac{\sum_{i=1}^n v_{c,i}\, w_{c,i}}{\sum_{i=1}^n w_{c,i}}, \quad (3)$$

where the subscript $c = \{z, m, v\}$ of the velocity planisphere $v_c$ refers to the three velocity components that we consider—zonal ($z$), meridional ($m$), or vertical ($v$)—and $w_c$ is the corresponding weight, defined as $w_c = 1/\sigma_c^2$, where $\sigma_c$ is the standard deviation of the photon noise. The subscript $i$ refers to the measurement number (see Appendix B for more details).

In practice, this step involves stacking the deprojected dopplergrams by positioning them in longitude according to Jupiter's reference rotation period. Thanks to the long duration of our observation campaign, every point on the three planispheres results from the average of many measurements taken at different times.

## 4. Results

### 4.1. Zonal Wind Profile

Figure 3 (left panel) displays the mean zonal wind profile as a function of latitude[19] from the complete 2018 JIVE data. The profile is compared with that from cloud tracking derived from the 2018 HST/OPAL data, which was then degraded to match the observing conditions of JIVE. In Figure 3 (right panel), we compare the actual HST profile with the JIVE profile to which we added the difference between the original HST profile minus the degraded one.

Both ways of comparing the HST cloud-tracking and JIVE Doppler profiles lead to a good agreement, which is in slight contrast with what Gonçalves et al. (2019) obtained from the 2015 and 2016 JOVIAL measurements. At the time, Gonçalves et al. (2019) reported a significant discrepancy between the JOVIAL and HST/OPAL profiles in the North Equatorial Belt and northern part of the Equatorial Zone (latitude ~5°). This discrepancy was thought to likely result from a purely instrumental and data processing bias, but a real physical explanation originating from a difference of sensitivity between both techniques could not be ruled out. Our new measurements tend to reinforce the bias hypothesis, even though a small difference is still present at the $1\sigma$ level. A possible explanation of the improved match between the Doppler and cloud-tracking results can be found in the improved data quality. In particular, the seeing never exceeded $2''$, whereas it was systematically larger than $2''\!.5$ during the JOVIAL observations obtained at the Calern observatory in 2015 and 2016. The values of the curves in Figure 3 can be found in Table 2 in Appendix C and are provided in machine-readable format.

---
[19] Latitudes are planetocentric unless otherwise indicated.

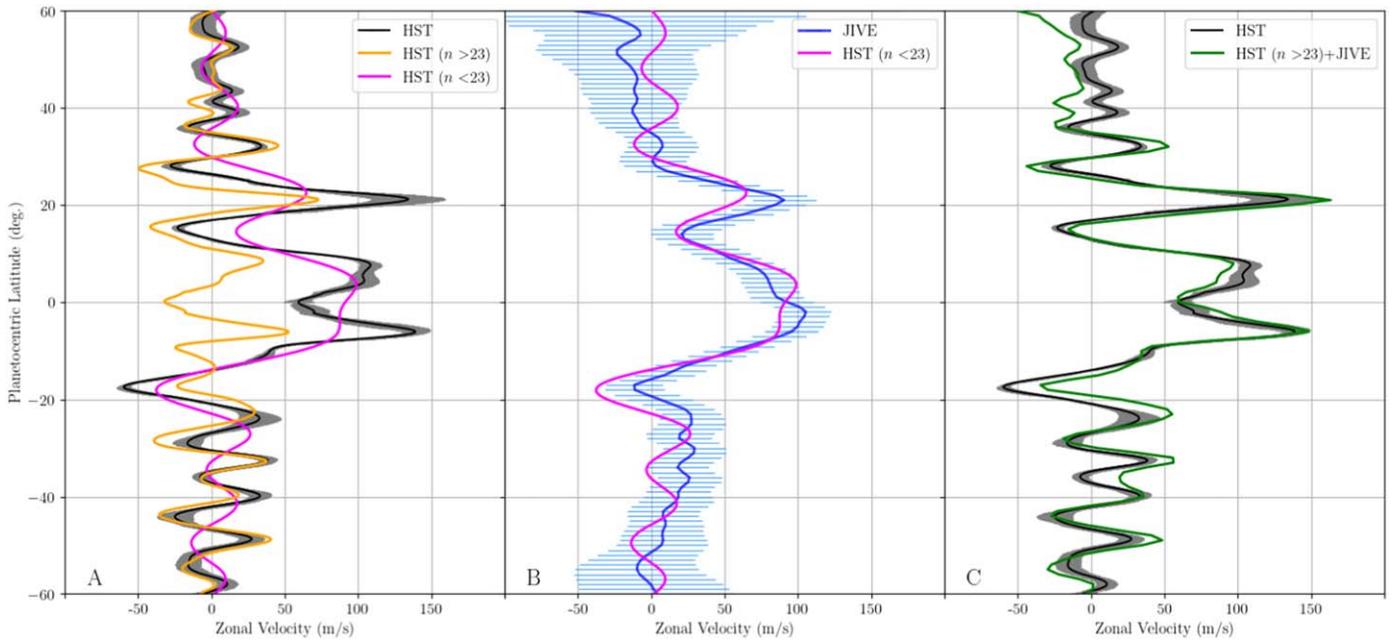

**Figure 4.** (A) Legendre decomposition of the HST profile for $n > 23$ (orange line) and $n < 23$ (magenta line). (B) Comparison between the JIVE profile and the $n < 23$ Legendre decomposition. (C) Comparison of the HST profile and the sum of the JOVIAL profile and the $n > 23$ Legendre decomposition.

We compare the JIVE profile with the energy-based decomposition of the zonal wind profile proposed by Galperin et al. (2001). In practice, they computed the zonal energy spectra of Jupiter (and Saturn) by decomposing the zonal profile obtained by cloud tracking into a set of Legendre polynomials, characterized by a wavenumber $n$, and interpreted the spectrum in terms of atmospheric regimes. In particular, the energy spectrum showed a clear change of slope at $n \approx 20$ by being almost flat under that limit and dropping down above it. The flatness of the spectrum in the low wavenumbers represents some form of friction at a large scale. According to Galperin et al. (2001), the low wavenumbers mainly represent the equatorial jet (see their Figure 3). They conjectured that these jets are a manifestation of the large-scale energy condensation due to inverse energy cascade under the influence of large-scale friction, where the higher modes, which account for the higher-frequency components of the wind profile, have a decaying spectrum characteristic of quasi-one-dimensional turbulence.

We repeated Galperin et al.'s (2001) Legendre decomposition on the 2018 HST profile and observed that up to $n = 23$; the difference between the lower mode decomposition and the new Doppler profile is minimized. This value of $n = 23$ roughly corresponds to the $n \approx 20$ of Galperin et al. (2001). Figure 4 compares our new Doppler wind profile with two profiles reconstructed from the decomposition in Legendre polynomials of the cloud-tracking wind profile, one corresponding to the first 23 modes and the other corresponding to the modes larger than 23. This indicates that our new Doppler profile mainly captures the signature of the equatorial jet in terms of the kinetic energy spectrum. Shaw et al. (2022) also compared their zonal wind profile obtained with the Doppler technique to the Legendre decomposition and showed that their profile was sensitive to the high-order components and was biased at low orders. The fact that our current observations are sensitive to orders lower than those retrieved by Shaw et al. (2022) would be the result of the difference in spatial resolution. With an improved seeing and spatial resolution, JIVE's observations would have allowed us to access higher orders of the Legendre decomposition.

### 4.2. Zonal Velocity Maps

Figure 5 displays 2D zonal velocity maps of Jupiter obtained from JIVE (top panel) and HST/OPAL (bottom panel) data. The general agreement is good if we keep in mind the difference in spatial resolution. The zonal velocity is mostly uniform as a function of longitude at a given latitude with the notable exception of the GRS, which appears clearly in the form of a shear near 20°S.

Weak longitudinal fluctuations of the zonal circulation are visible in several places. In the north, we notice variations in the region between 15° and 20°, where a train of large anticyclones was present in 2018 (Simon et al. 2018). In the south, we notice a smooth pseudoperiodic variation between −20° and −40° where large anticyclones, such as the Oval BA at about −30° and a train of them at −40°, are present. These longitudinal variations might also result from the incomplete time coverage of the observations, as each night does not cover a complete rotation of Jupiter, so a given region may have been observed in less time or with different observing conditions.

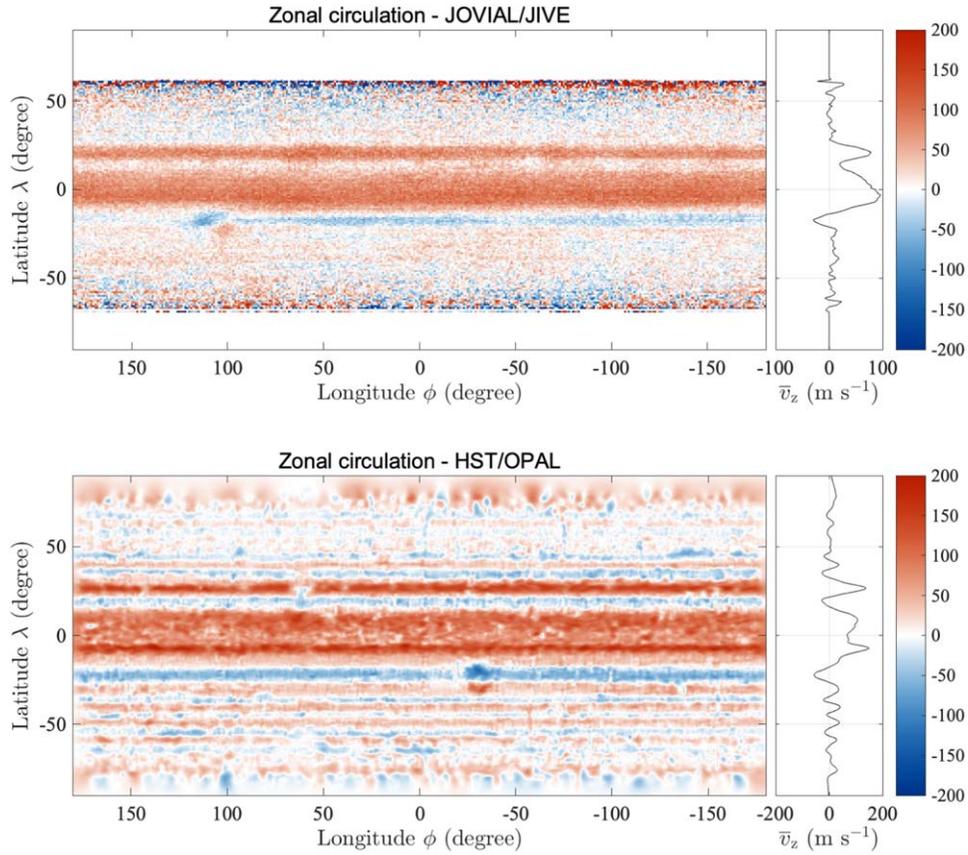

**Figure 5.** Jupiter zonal circulation from Doppler spectroscopy and cloud tracking. Top panel: deprojected velocity maps of Jupiter based on the 2018 JIVE campaign at DST. Bottom: zonal velocity map retrieved from cloud tracking based on HST OPAL data. Color scales indicate the velocity expressed in m s$^{-1}$. Longitudes and latitudes are planetocentric. Positive zonal velocity is prograde, that is, eastward.

Actually, much of the possible longitudinal variations of the zonal circulation may have been partially erased by the way we assemble the deprojected dopplergrams into planispheres. Indeed, the projection of the Doppler measurements onto a planisphere is performed by assuming a solid rotation rate. As we average over 20 days of data, the differential rotation washes out most of the possible longitudinal variations of the zonal circulation, especially in the regions with a strong prograde or retrograde wind. The way we assembled the maps preserved most of the GRS, though, because the position of the GRS is almost fixed in the rotation referential that we used (system III). Another averaging could have been done by assuming that the structures follow the wind at that latitude. It would give more details in the area of the northern equatorial belt, where many dynamical structures are present. However, doing so is a complex task to perform given the long duration of our observations. Indeed, applying such a shift based on the mean zonal profile for almost 1 month would spread the GRS over a large range of longitude.

Finally, by looking closely at the GRS in both the JIVE and HST zonal planispheres, we note that the positive and negative parts of the GRS are not on top of each other in the JIVE data, contrary to the HST map. This apparent asymmetry of the GRS in the JIVE map could be explained by improper modeling of the dynamics around the GRS when taking care of the velocity bias (Section 3.2). Indeed, the strong flux gradient combined with the fast rotation and the GRS's drift relative to the winds produces a spurious shift in the Doppler measurement, which could only be removed by complete modeling of the velocity, including the meridional motion.

### 4.3. Meridional and Vertical Velocity Maps

As anticipated in Section 3.2, dealing with meridional and vertical circulations is not as straightforward as for zonal winds since the contamination of the Doppler signal by instrumental effects is not averaged out by the rotation of the planet. In addition, decomposing the dopplergrams into meridional and vertical maps is a degenerate problem. The vertical component and the meridional are fully correlated because their projection factors only differ by a factor of $\tan \lambda$, where $\lambda$ is the latitude (Appendix B). Therefore, it is impossible to completely disentangle them, since all of our measurements are obtained from a single location, which is near the equatorial plane of Jupiter. We nevertheless represent in Figure 6 the 2D maps of the meridional and vertical velocities as found by assuming that the other component is null. The values thus obtained are upper limits. We also stress that our measurements are not absolute, as the mean Doppler value at the surface was arbitrarily set to 0. This comes out from our data processing. In any case, what matters most are the variations between the different regions.

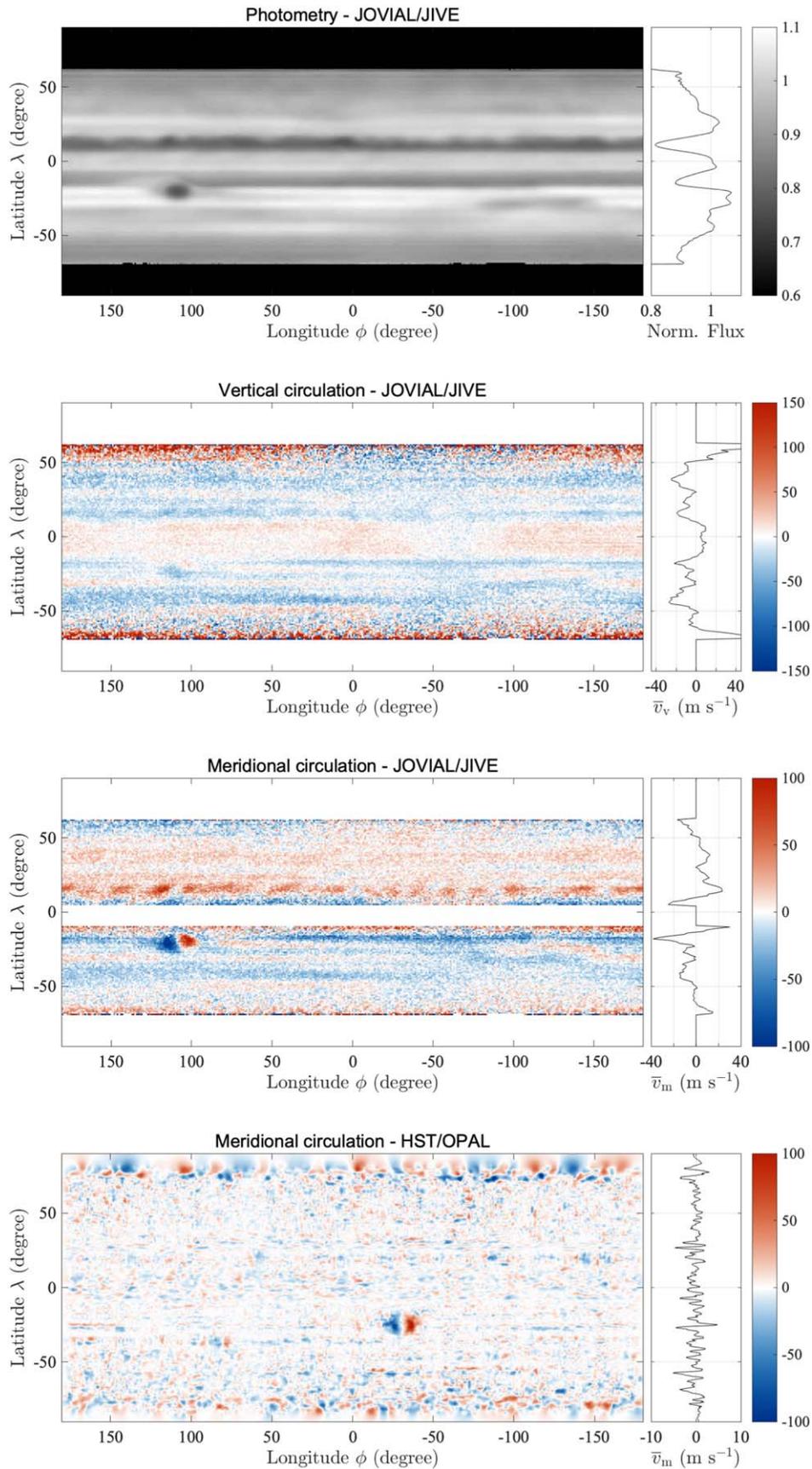

**Figure 6.** Top panel: photometric map of Jupiter from the JIVE campaign. Middle panels: inferred vertical (top) and meridional (bottom) from deprojected dopplergrams (in m s$^{-1}$) derived from the 2018 JIVE campaign at DST. Vertical (resp. meridional) velocities are obtained by assuming that meridional (resp. vertical) velocities are null. The deprojected meridional velocity field is discarded in the equatorial region to avoid the plot being dominated by noise. Bottom panel: meridional velocity field component (in m s$^{-1}$) derived from HST observations. Positive meridional velocity is northward, and positive vertical velocity is upward.

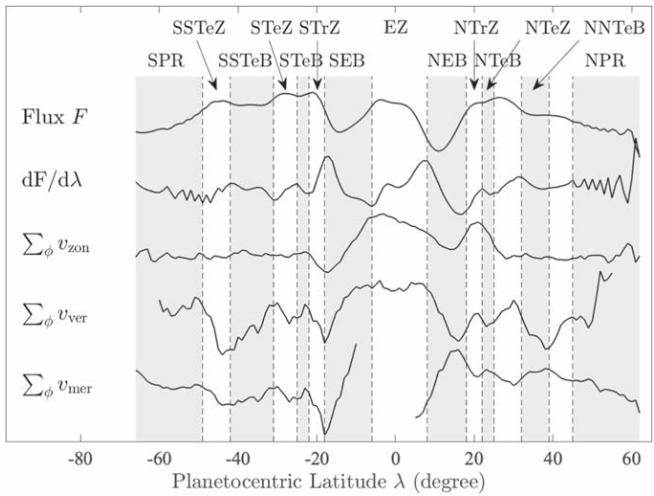

**Figure 7.** Comparison between the photometric and inferred velocity profiles from the JIVE data. The flux $F(\lambda)$ is the photometric map from Figure 6 (top panel) averaged over the longitudes. Its derivative $\partial F/\partial \lambda$ is displayed as the second line. The three bottom lines show the same for the projected zonal, vertical, and meridional maps displayed in Figure 6. The background gray and white areas indicate the location of the belts (gray) and zones (white). From left to right, the acronyms correspond to the usual nomenclature: SPR for South Polar Region, SSTe (Z or B) for South-South Temperate (Zone or Belt), STe (Z or B) for South Temperate (Zone or Belt), STr (Z or B) for South Tropical (Zone or Belt), SEB for South Equatorial Belt, and EZ for Equatorial Zone. The nomenclature is the same in the northern hemisphere, with N in place of S.

Let us first focus on the map of vertical velocities, which is expected to dominate the deprojected composite vertical and meridional Doppler signal in the equatorial region. We observe a uniform vertical motion over the Equatorial Zone that drops by about $20 \text{ m s}^{-1}$ at the northern end of the North Equatorial Belt and the southern end of the South Equatorial Belt. We notice a slight enhancement of the upward motion at the edge between the zone and the belt. The vertical velocity in the Equatorial Zone appears to be quite uniform, with longitudinal variations that, as for the zonal flows, are probably the results of an uneven distribution of observations in terms of Jovian longitudes. We do not attempt to interpret the vertical velocities at higher latitudes as we expect the values to be largely spurious, the signal being dominated by the meridional component there.

Qualitatively, the vertical circulation map is compatible with the model proposed by Duer et al. (2021), with a slightly upward motion in the Equatorial Zone and a downward motion at the edge of that zone and in the adjacent belts. However, the vertical velocities appear to be larger than expected. Mixing-length theory yields an average velocity of $1 \text{ m s}^{-1}$ around the 1 bar pressure level in order to transport Jupiter's intrinsic luminosities by small-scale convective motions (Guillot et al. 2004). For mesoscales ($\approx 10^3$ km), the estimates from cloud-ensemble models indicate velocities in quiet regions that are of the order of a few $\text{m s}^{-1}$ (e.g., Sugiyama et al. 2014). For synoptic scales ($\approx 10^4$ km), Read et al. (2005) estimated the vertical velocities to be of the order of a few $\text{cm s}^{-1}$.

As described in the previous sections, photometric inhomogeneities coupled with Jupiter's fast rotation cause biases in the dopplergrams. It is tantalizing to attribute most of the content of the vertical map to biases because of the unexpectedly large values that we measure. In such a case, the inferred vertical velocity should be correlated to the photometric map or its derivative. To appreciate whether this is the case, Figure 7 compares the projection of the three 2D maps along the longitude $\phi$ onto the latitude $\lambda$ with the photometric profile $\sum_\phi F(\lambda, \phi)$ and its derivative. If a correlation between the vertical profile $\sum_\phi v_{\text{ver}}(\lambda, \phi)$ and $\partial(\sum_\phi F(\lambda, \phi))/\partial \lambda$ seems plausible between 10°N and 40°N, nothing similar is visible in the southern hemisphere. We even see an anticorrelation between the two at $-20°$. We cannot identify a correlation between $v_{\text{ver}}$ and $F$ or $\partial F/\partial \lambda$. Therefore, from a theoretical point of view, there is no obvious evidence of observational or instrumental bias in the inferred vertical velocity map.

Let us comment on the meridional map. We first point out that in Figure 6, we discarded the region from 9°S to 4°N. Indeed, the meridional projection factor is 0 at a latitude located halfway between the subsolar and the subterrestrial latitudes, implying that any inferred meridional velocity in this region is not realistic.

The overall aspect of the 2D map appears to be very complex and in strong disagreement with the meridional map obtained by cloud tracking from the HST/OPAL data (Figure 6). In particular, we note banded structures (northward or southward) that are essentially functions of the latitude only. This type of signal is totally absent in the meridional map from HST data. Again, given how discrepant the two maps are, and given that meridional and vertical motions are more sensitive to instrumental biases, it is natural to think that most of the features in the meridional map are not real. However, as for the inferred vertical velocity map, there is no evidence that these maps are dominated by biases.

To clarify our ability to measure meridional fields without suspicion of complex instrumental biases, we then focused on the only region that is larger than the typical PSF, mostly photometrically homogeneous, and where we expect a clear rotating signal: the GRS. In Figure 8, we show a zoom of the meridional maps obtained by JIVE and HST. Once corrected from the bias at the edges of the GRS due to the photometric gradient and the strong rotation velocity, the map shows a meridional motion at the eastern and western sides of the GRS, which is in good agreement with the known rotation velocity of the GRS. The maximum velocity is of the order of $100 \text{ m s}^{-1}$, which is similar to what was reported by Wong et al. (2021). We note that the shape of the Doppler meridional velocity field appears asymmetrical, as is the case for the zonal motion. A crosstalk in the data processing between zonal and meridional motion likely explains this shape, as the model used for the correction of the Doppler measurement includes only a zonal wind component, based on the cloud-tracking profile and constant along each latitude. This is certainly not valid in the case of the latitude of the GRS, where the local dynamics on either side of the GRS are very different. This could induce an uncorrected bias of a few tens of $\text{m s}^{-1}$ in estimating the wind near the GRS.

### 4.4. Reflectivity versus Divergence of Horizontal Flow

The last test we perform with the data consists of comparing the divergence of the horizontal velocity field inferred from the JIVE data with the reflectivity—photometry minus limb-darkening model—of the planet. According to the conservation of mass, we have

$$\frac{1}{\rho}\frac{d\rho}{dt} + \boldsymbol{\nabla} \cdot \boldsymbol{u} = 0, \quad (4)$$

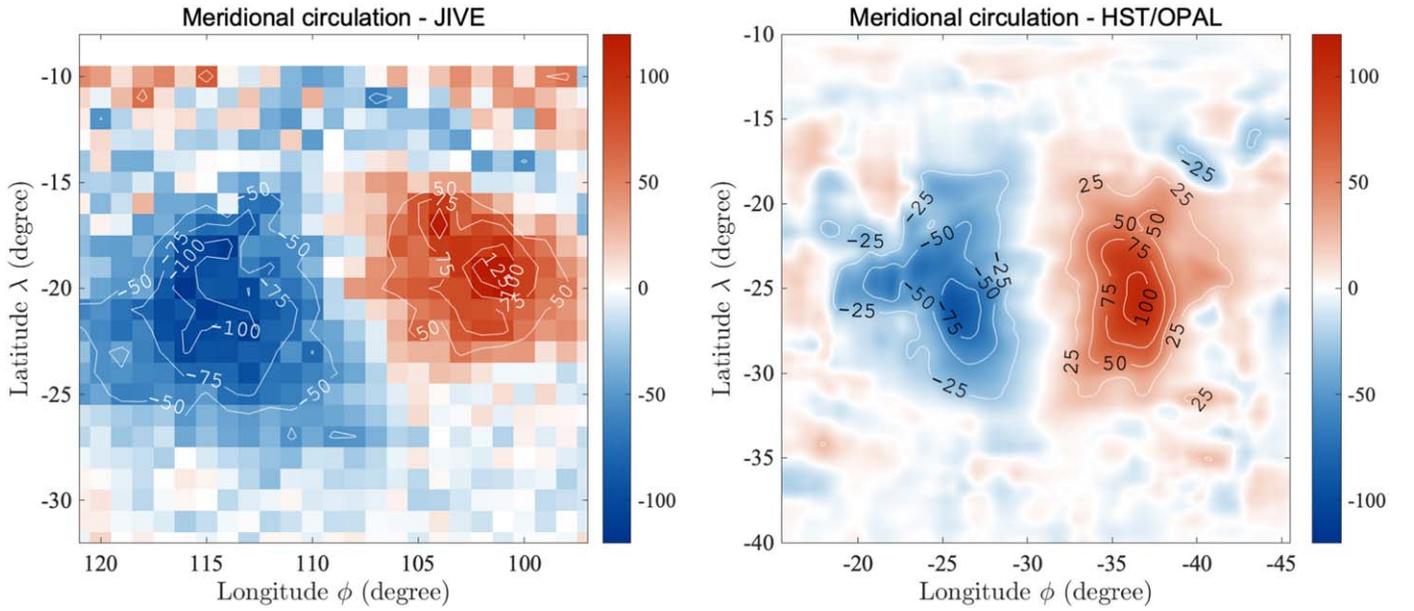

**Figure 8.** Comparison of the meridional circulation of the GRS from JIVE (left) and HST/OPAL (right). Velocities are expressed in m s$^{-1}$.

where $\rho$ is the density and $\boldsymbol{u} = \boldsymbol{v}_{\mathrm{zon}} + \boldsymbol{v}_{\mathrm{mer}} + \boldsymbol{v}_{\mathrm{ver}}$ is the atmospheric velocity field. By assuming that $\rho$ does not change significantly during the observations at the altitude where the velocities are measured, we can conclude that $\nabla \cdot \boldsymbol{u} = 0$. By separating the horizontal and vertical components into $\boldsymbol{u}_h = \boldsymbol{v}_{\mathrm{zon}} + \boldsymbol{v}_{\mathrm{mer}}$ and $w = \boldsymbol{v}_{\mathrm{ver}}$, we have

$$\frac{\partial w}{\partial z} = -\nabla_h \cdot \boldsymbol{u}_h. \quad (5)$$

Therefore, a positive (negative) divergence of the horizontal wind field would correspond to a decrease (increase) in vertical velocity as a function of height. In the hypothesis of a null vertical velocity at the tropopause, a positive divergence implies a positive velocity (upflow). However, our vertical velocity profile does not fully coincide with that, in particular in the region of the equator between 5° and 20°N and S. It might be the signature of the existence of wave activity in these regions. In any case, a correlation between the reflectivity profile and the horizontal divergence profile (Figure 9) is not surprising, as shown by Fletcher et al. (2021, Figure 12(a) and references therein).

## 5. Discussion and Prospects

### 5.1. Technical Point of View

In this paper, we report the first-ever maps of the atmospheric circulation of any planet obtained by Doppler spectroscopy in the visible domain. We generated 3D velocity maps of Jupiter by combining dopplergrams obtained from several complete rotations of the planet. It is a premiere for the zonal wind map alone, as well as for the meridional and vertical components. In general, we demonstrate that Doppler imaging is an actual option for studying the atmospheric dynamics of the planets of the solar system. In particular, it opens a unique way of investigating vertical flows, which cannot be accessed with cloud-tracking methods.

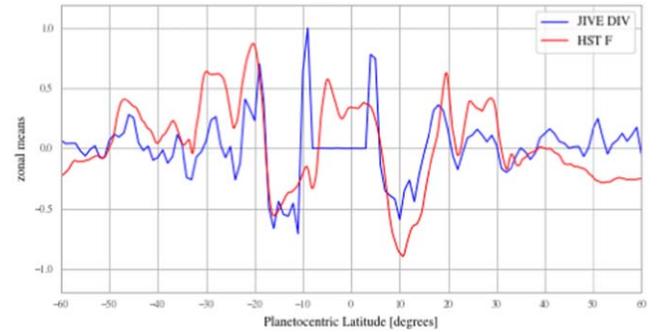

**Figure 9.** Normalized zonal reflectivity profile corresponding to the 502 nm OPAL map (red) compared to the normalized profile of $-\nabla_h \cdot \boldsymbol{u}_h$ from the JIVE data (blue).

It is important to acknowledge and summarize the main technical difficulties that we encountered in this work. From a purely instrumental point of view, there exist biases caused by an inhomogeneous response of the instrument, which caused fixed line-of-sight velocity patterns in the field of view. The way we found to circumvent this difficulty consisted of observing Jupiter with different orientations in the field, thanks to the telescope configuration, to average those effects out. We should also apply this method for the other sites of the JOVIAL network. It is already possible for the observations in Calern, where we added a derotator to the instrument. It will also be needed at the Okayama site. From an observational point of view, our main limitation is the degradation of the image by atmospheric seeing, whose main consequence is to blend regions of different photometry and line-of-sight velocity together. The correction for the effect of seeing is based on the zonal motion, including the fast rotation and the zonal wind profile, for which we have a good model. There is no equivalent for meridional and vertical motions, so their effects on the measurement are not taken into account for the velocity correction. It has to be noticed, however, that these velocities and their gradient are small as compared to the fast rotation, so the effect should be negligible. Vertical and meridional velocity

The Planetary Science Journal, 5:100 (17pp), 2024 April                                                                                                           Schmider et al.fields are more difficult to extract from the dopplergrams than the zonal one, for several reasons. First, the velocities are supposed to be smaller than the zonal winds and, in any case, much smaller than the fast rotation velocity. Second, the meridional and vertical projection factors have the same sign for all the longitudes on Jupiter, so the rotation of the planet will not suppress any spurious velocity signal, contrary to the zonal measurements. Finally, vertical and meridional velocities cannot be determined independently, as all the observations were obtained from the Earth at a constant latitude on Jupiter, so we cannot attribute the Doppler contribution to one or the other component.

That being said, the methods we developed are robust, and the limitations that we face would not be met in the absence of seeing alteration, such as from space.

### 5.2. Scientific Point of View

From a scientific point of view, we must point out the zonal wind profile that we get and the zonal map, which can be considered as the first successful effort to measure the velocity field at the surface of Jupiter from Doppler measurements obtained from the ground. As we observe in the visible, at a wavelength of 519 nm, we clearly see the top of the cloud cover at about 1 bar. We also want to point out the noise level that was reached. The theoretical noise level in the zonal wind profile is of the order of $1\ \mathrm{m s}^{-1}$. The photon noise level seen on each image is close to the theoretical noise. Therefore, the larger dispersion of the measurements used to produce the zonal profile, and the planisphere could only result from actual variations of the zonal wind along the longitude and during the observing period.

We cannot guarantee the reliability of the new findings that appear in our zonal wind profile and in our 2D maps, as some unknown biases may exist in the velocity data. However, we think it is important to keep in mind a couple of features that deserve to be followed up in future measurements, with or without the same instrumentation.

First, the discrepancy between the zonal wind profile that was reported by Gonçalves et al. (2019) is still present even though it is less prominent. We can still assume that it originates from some form of instrumental bias, but it is puzzling to realize that it appears at the same latitude on Jupiter, whereas the telescope is different and the seeing conditions are much different. The observing procedure used, by rotating Jupiter in the field regularly, should prevent any remaining instrumental bias.

It is legitimate to question whether this difference arises from the difference between the cloud tracking and the direct wind measurement Doppler technique. One possibility would be wave activity affecting cloud tracking and Doppler measurement in a different manner.

Indeed, the most surprising feature in the meridional velocity is a strong short-period modulation of the northern equatorial belt, where "hot spots," or cloud clearings, are known to exist. An alternation of positive and negative meridional circulation, relative to the mean level at that latitude, could indicate a physical phenomenon. It is known that Rossby waves are present in this region (e.g., Giles et al. 2019), and the large cloud plumes show evidence of rotation in Galileo and Voyager data (Vasavada et al. 1998; Choi et al. 2013). Cloud tracking reflects the motion of cloud structures that are by definition driven by the region of a given pressure and relative humidity level. We may conjecture that in this specific latitude range, the Doppler technique is able to measure motions related to the waves that are not visible in terms of cloud pressure-driven structures.

Finally, it has to be noticed that an underlying hypothesis for the calculation of the planispheres is that the velocity field remains constant with time at each point of the surface all along the observations. Nevertheless, all measurements have been done on the Jovian dayside, and we never see what happens on the nightside, even if we believe that it might not be very different.

### 5.3. Future Plans

The reliability of the results can be enhanced through further advancements in both observational methods and data processing techniques. We have suggested potential avenues for improvement, such as refining the estimation of the PSF during further observations and conducting simultaneous inversions of the velocity field parameters. Additionally, enhancing the data quality, for instance, by utilizing an adaptive optics system simultaneously, would lead to more accurate results. We are presently developing one at the Calern observatory for that purpose (Buralli et al. 2022). However, the correction of the atmospheric turbulence can only be partial, as Jupiter, like the other planets of the solar system, is larger than the isoplanetic angle (typically 2″ at standard sites) on which the wave-front error remains coherent. Moreover, such a correction will remain very difficult in the visible domain.

True improvements would only come from space-based Doppler observations. In that case, the PSF would be known, limited only by diffraction, and it would be constant over time, allowing for a full inversion of the data. Moreover, measurements taken at different longitudes and latitudes would distinguish between the wind components and would provide much more precise data, enabling a comprehensive understanding of atmospheric dynamics. In line with the Uranus mission, ranked as a top priority in the NASA Decadal Survey, a Doppler imaging instrument with a focus on studying atmospheric dynamics and potentially detecting oscillation frequencies would be invaluable. Furthermore, a Jupiter observatory located at the L1 Lagrange point, as proposed by Hsu et al. (2019), would be an ideal instrument for studying the atmospheric dynamics of the giant planet. The spacecraft's orbital motion around the L1 point allows for velocity recovery from different lines of sight, and within 2 yr, we could achieve a noise level below $1\ \mathrm{mm\ s}^{-1}$ for oscillation measurements and wind measurements with a precision of the order of $0.1\ \mathrm{m\ s}^{-1}$.

### Acknowledgments

This work was possible thanks to the ANR JOVIAL and the JIVE in NM NASA EPSCoR program. The NMSU team would also like to acknowledge support from NASA Solar System Observations grant No. 80NSSC20K0672. This work used data acquired from the NASA/ESA HST Space Telescope, associated with the OPAL program (PI: Simon; GO13937) and archived by the Space Telescope Science Institute, which is operated by the Association of Universities for Research in Astronomy, Inc., under NASA contract NAS 5-26555. All maps are available at 10.17909/T9G593. P.G. wishes to thank Paola Gaza for her support while working on the manuscript.12

# Appendix A
# Extracting the Velocity Field

As described in Section 3.2, extracting the actual velocity field from the Doppler measurement entails having an estimate of the image flux $F$. Unfortunately, the measurements only give access to the flux map convolved with the PSF: $F_m = F * P$. Here, we demonstrate how to significantly reduce the biases introduced in the Doppler velocity maps by the PSF with the sole knowledge of $F_m$ and a theoretical model of $v_d$. Figure 10 shows the effect of the PSF on the images.

According to Civeit et al. (2005), at a given point of coordinates $(x, y)$ on the image, the measured line-of-sight velocity $v_m$ is expressed as:

$$v_m(x, y) = \frac{(F\, v_d * P)(x, y)}{(F * P)(x, y)} \quad (A1)$$

$$= \frac{\iint_{-\infty}^{+\infty} F(u, v) v_d(u, v) P(x - u, y - v)\, du\, dv}{F_m(x, y)}. \quad (A2)$$

Locally, the product of convolution is simply the integral of the photometry multiplied by the PSF $P$,

$$v_m(x, y) = \frac{\iint_{-\infty}^{+\infty} F(u, v) v_d(u, v) P(u', v')\, du'\, dv'}{F_m(x, y)}, \quad (A3)$$

where the relative coordinate system with respect to $(x, y)$ is rewritten as $u' = u - x$ and $v' = v - y$.

To simplify the convolution equation, we approximate both the photometry $F$ and velocity fields $v_d$ in the form of a first-order Taylor expansion. Around $(x, y)$, the local approximated photometry at the coordinates $(u, v)$ is

$$F(u, v) = F(x, y) + \frac{\partial F(x, y)}{\partial x} u' + \frac{\partial F(x, y)}{\partial y} v', \quad (A4)$$

and the velocity map is

$$v_d(u, v) = v_d(x, y) + \frac{\partial v_d(x, y)}{\partial x} u' + \frac{\partial v_d(x, y)}{\partial y} v'. \quad (A5)$$

In the numerator of Equation (A3), we have at first order

$$F(u, v) v_d(u, v) = \left[ F(x, y) + \frac{\partial F(x, y)}{\partial x} u' + \frac{\partial F(x, y)}{\partial y} v' \right]$$
$$\times \left[ v_d(x, y) + \frac{\partial v_d(x, y)}{\partial x} u' + \frac{\partial v_d(x, y)}{\partial y} v' \right]. \quad (A6)$$

To compact the writing, we drop $(x, y)$. Then,

$$F(u, v) v_d(u, v) = F v_d + F \frac{\partial v_d}{\partial x} u' + F \frac{\partial v_d}{\partial y} v'$$
$$+ v_d \frac{\partial F}{\partial x} u' + v_d \frac{\partial F}{\partial y} v' + \ldots$$
$$+ \frac{\partial F}{\partial x} \frac{\partial v_d}{\partial x} u'^2 + \frac{\partial F}{\partial y} \frac{\partial v_d}{\partial y} v'^2$$
$$+ \frac{\partial F}{\partial x} \frac{\partial v_d}{\partial y} u' v' + \frac{\partial F}{\partial y} \frac{\partial v_d}{\partial x} u' v'. \quad (A7)$$

The integrals of odd functions such as the terms that include $u'$, $v'$, or their product are null. Therefore, we are left with the terms $F v_d$, $\frac{\partial F}{\partial x} \frac{\partial v_d}{\partial x} u'^2$, and $\frac{\partial F}{\partial y} \frac{\partial v_d}{\partial y} v'^2$.

From Equation (A3), we have

$$v_m(x, y) = \frac{1}{F_m(x, y)} \iint_{-\infty}^{+\infty} F(x, y) v_d(x, y) P(u', v')\, du'\, dv'$$
$$+ \frac{1}{F_m(x, y)} \iint_{-\infty}^{+\infty} \frac{\partial v_d(x, y)}{\partial x} u'^2 \frac{\partial F(x, y)}{\partial x}$$
$$\times P(u', v')\, du'\, dv'$$
$$+ \frac{1}{F_m(x, y)} \iint_{-\infty}^{+\infty} \frac{\partial v_d(x, y)}{\partial y} v'^2 \frac{\partial F(x, y)}{\partial y}$$
$$\times P(u', v')\, du'\, dv'. \quad (A8)$$

Let us consider the three integrals individually by letting $1/F_m$ aside for now. The first term in Equation (A8) can be written

$$\iint_{-\infty}^{+\infty} F(x, y) v_d(x, y) P(u', v')\, du'\, dv'$$
$$= F(x, y) v_d(x, y) \iint_{-\infty}^{+\infty} P(u', v')\, du'\, dv' \quad (A9)$$

$$= F(x, y) v_d(x, y), \quad (A10)$$

by assuming the PSF to be a Gaussian function of standard deviation $\sigma_P$ normalized such as its integral is one:

$$P(u', v') = \frac{1}{2\pi \sigma_P^2} e^{-\frac{u'^2 + v'^2}{2\sigma_P^2}}. \quad (A11)$$

Regarding the second term (and similarly for the third term), we have

$$\iint_{-\infty}^{+\infty} u'^2 \frac{\partial F(x, y)}{\partial x} \frac{\partial v_d(x, y)}{\partial x} P(u', v')\, du'\, dv'$$
$$= \frac{\partial F(x, y)}{\partial x} \frac{\partial v_d(x, y)}{\partial x} \iint_{-\infty}^{+\infty} u'^2 P(u', v')\, du'\, dv' \quad (A12)$$

$$= \sigma_P^2 \frac{\partial F(x, y)}{\partial x} \frac{\partial v_d(x, y)}{\partial x}. \quad (A13)$$

Hence, Equation (A8) becomes

$$v_m(x, y) = v_d(x, y) \frac{F(x, y)}{F_m(x, y)} + \frac{\sigma_P^2}{F_m(x, y)}$$
$$\times \left( \frac{\partial F(x, y)}{\partial x} \frac{\partial v_d(x, y)}{\partial x} + \frac{\partial F(x, y)}{\partial y} \frac{\partial v_d(x, y)}{\partial y} \right). \quad (A14)$$

Ideally, from Equation (A14), we could extract $v_d$ from $v_m$ if we had $F$ and the partial derivatives of $F$ and $v_d$ with respect to $x$ and $y$. The approximation we employ in this work consists of three aspects. First, we assume that the flux is not too altered by the seeing so that $F/F_m \approx 1$. Second, we are assuming that the derivative of the flux with respect to $x$ and $y$ is not altered by the seeing, meaning $\partial F/\partial x \approx \partial F_m/\partial x$ and $\partial F/\partial y \approx \partial F_m/\partial y$. In other words, it assumes that the gradient of the flux will be smoothed by the PSF in the same way as the flux itself. Third, since the variations of the line-of-sight velocities are smooth, we employ a theoretical value of $\partial v_d/\partial x$ extracted from the HST/OPAL zonal wind profile. Hence, our estimate $\widehat{v_d}$ of the line-of-sight velocity field $v_d$ that takes into account the degradation of the dopplergram by the PSF is

$$\widehat{v_d}(x, y) \approx v_m(x, y) - \frac{\sigma_P^2}{F_m(x, y)} \times \left( \frac{\partial F_m(x, y)}{\partial x} \frac{\partial v_{z,\text{HST}}(x, y)}{\partial x} + \frac{\partial F_m(x, y)}{\partial y} \frac{\partial v_{z,\text{HST}}(x, y)}{\partial y} \right), \quad \text{(A15)}$$

where $v_{z,\text{HST}}$ is the zonal wind value obtained from the HST data.

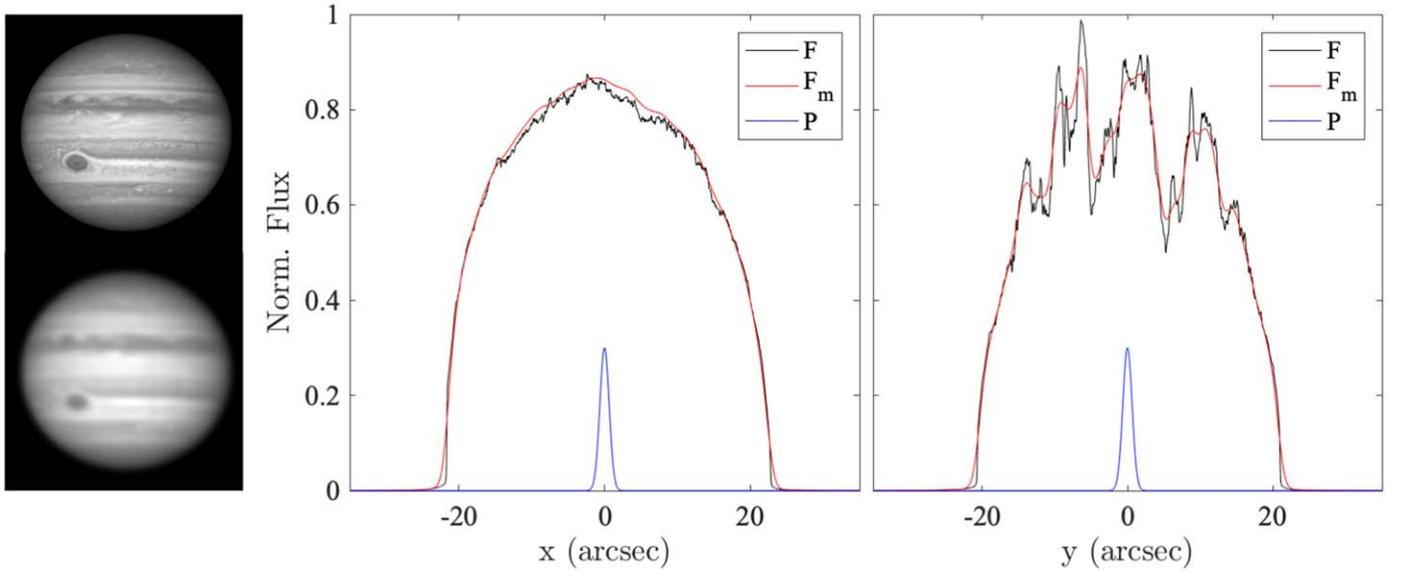

**Figure 10.** Effect of the PSF on an image of Jupiter taken by HST. The top left image was taken by HST on 2018 April 17 and is used as $F(x, y)$. The bottom left image is the same but convolved by a Gaussian PSF corresponding to a seeing of $1.''5$; it corresponds to $F_m(x, y)$. The middle panel shows a cut of $F$ (black) and $F_m$ (red) along the $x$-axis in the equatorial region, while the right panel shows a cut along the $y$-axis at the central meridian. The PSF $P$ is represented under the photometric profiles as a blue line. The HST image has the observation ID idg216i9q and is available on the MAST. It was rectified from the $133°.2508$ inclination on the detector with a bilinear interpolation.

To quantify what error is made with such an approximation, we ran a simulation based on a high-resolution image of Jupiter taken by HST and a simple solid-body rotational model $v_d$. We computed the associated dopplergram $v_m$ by applying Equation (A2) and then extracted the estimate $\widehat{v_d}$ of $v_d$ by applying Equation (A15). Figure 11 shows the maps of the differences $v_m - v_d$ (left panel) and $v_m - \widehat{v_d}$ (middle) and the difference between the two $\widehat{v_d} - v_d$ (right). The two maps are in good agreement—the true and retrieved dopplergrams differ by a few m s$^{-1}$—except for the very edges. Indeed, Equation (A15) cannot work properly at the edge, because the photometric gradient is maximum, and because the PSF blends regions inside and outside Jupiter. To circumvent the edge issue, the data processing pipeline first flattens the dopplergrams by subtracting the theoretical map of a Lambert spheroid with the solid-body rotation of Jupiter (system III) before applying Equation (A15). Once cleared of the combined effect of the steep photometric gradient and the large solid-body rotation, the first-order development should work fine on the whole planetary disk.

However, in practice, we observed a remnant bias in the dopplergrams, which we could minimize by multiplying $\sigma_P$ by a factor of between 1 and 2, depending on the seeing. The actual correction of the photometric bias was ultimately performed with the following expression:

$$\widehat{v_d}(x, y) \approx v_m(x, y) - \xi \frac{\sigma_P^2}{F_m(x, y)} \times \left( \frac{\partial F_m(x, y)}{\partial x} \frac{\partial v_{z,\text{HST}}(x, y)}{\partial x} + \frac{\partial F_m(x, y)}{\partial y} \frac{\partial v_{z,\text{HST}}(x, y)}{\partial y} \right), \quad \text{(A16)}$$

with $\xi \approx 1.5$. If the reasons for needing such a factor are not fully clear, our understanding is that the seeing estimate is systematically underestimated. Inaccuracies about the seeing arise from the way we measure it—based on Jupiter's size on the detector instead of using a reference star in the field of view. An additional issue is that we assumed the PSF to be a Gaussian function. This assumption has the great advantage of leading Equation (A8) to an analytical solution, but real PSFs may significantly differ from a Gaussian model, since they result from the combination of optical diffraction and atmospheric turbulence that is partially corrected by tip-tilt or adaptive optics mechanisms. In a general case, PSFs are modeled by a Moffat function, whose parameters depend on the atmospheric conditions (Fusco et al. 2020). In other words, the $\xi$ factor addresses the departure to an approximate description of the PSF.

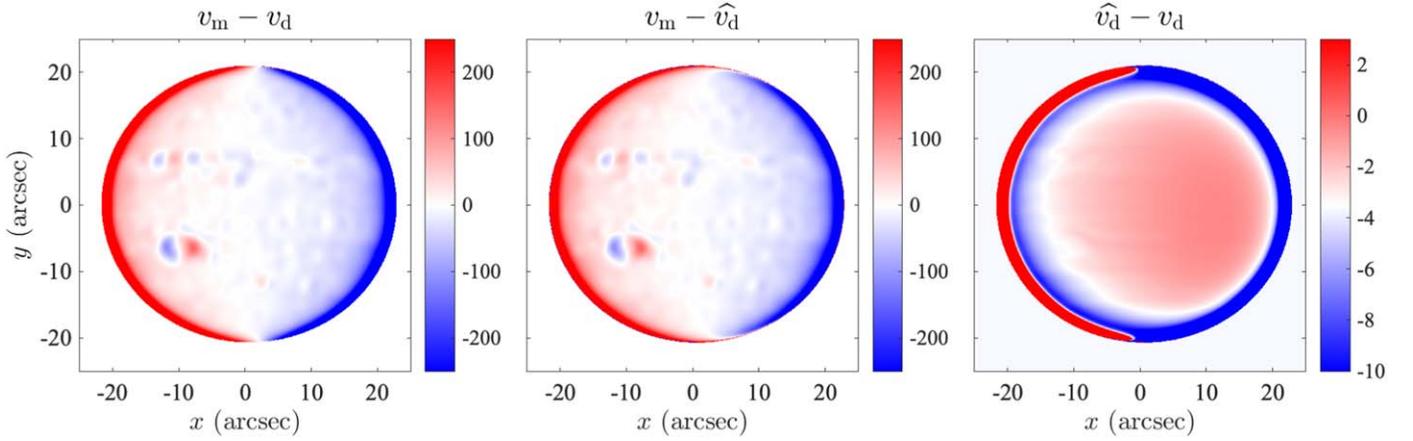

**Figure 11.** Bias introduced by the PSF on the dopplergrams and its suppression thanks to the first-order developments of $F_m$ and $v_d$ (Equations (A4) and (A5)). Left panel: simulation of the difference between the measured and true velocity maps $v_m - v_d$, i.e., the velocity bias, introduced by a 2″ Gaussian PSF. The image used in this simulation is the same as with Figure 10. For the velocity map, we employed a solid-body rotator with Jupiter's period measured in the system III. Middle panel: difference between the measured data and the estimate of the true dopplergram $\widehat{v_d}$ obtained by applying Equation (A15). Right panel: difference between the true and retrieved velocity maps $\widehat{v_d} - v_d$. Color bars indicate line-of-sight velocities expressed in m s$^{-1}$.

## Appendix B
## From Velocity Maps to Planispheres

The zonal velocity maps can be obtained by dividing the dopplergrams by the zonal projection factor $f_{zon}$ (Gaulme et al. 2018 and Figure 12):

$$v_{zon} = \frac{v_m}{|f_{zon}(x, y)|} C_{zon}. \quad (B1)$$

This is not exact, as $v_m$ is the sum of all of the components of the atmospheric circulation, and not only the zonal one. However, as we show next, we average many images taken at different orientations of Jupiter. Therefore, any other component, if constant at a given point of the surface during one rotation of Jupiter, will cancel out because of the antisymmetric behavior of the zonal projection factor.

Since this projection factor approaches 0 near the meridian, the value tends toward infinity in this region. Therefore, before summing all the projected zonal velocity maps onto a planisphere, we need to apply a weighting equal to the inverse of the zonal velocity noise.

We can associate each measured radial velocity map with a theoretical noise map and derive a noise map associated with the zonal velocity through deprojection:

$$\sigma_{v_{zon}} = \frac{\sigma_{v_m}}{f_{zon}(x, y)} C_{zon}. \quad (B2)$$

By assuming that the zonal velocity remains locally constant at a given longitude during the observation period, it is possible to create an average zonal velocity map by computing the weighted average of all of the projections onto a planisphere with latitude $\lambda$ and longitude $\phi$ grids. Each projection of the radial velocity maps is adjusted in longitude according to the current value of the sub-Earth point longitude provided by the ephemeris. Each point on the average map is then calculated using the following weighted sum:

$$\overline{v_{zon}}(\lambda, \phi) = \frac{\sum_{i=1}^{N} v_{zon,i}(\lambda, \phi) \, w_{zon,i}(\lambda, \phi)}{\sum_{i=1}^{N} w_{zon,i}(\lambda, \phi)} \quad (B3)$$

with

$$w_{zon,i}(\lambda, \phi) = \frac{1}{\sigma^2_{v_{zon,i}}(\lambda, \phi)} \quad (B4)$$

and where $v_{zon,i}(\lambda, \phi)$ is the estimated zonal velocity map of the image $i$ and $\sigma_{v_{zon,i}}$ is the deviation map of its associated noise map.

The zonal velocity $v_{zon,i}(\lambda, \phi)$ is obtained from the image $v_{zon,i}(x, y)$ using the relationship that connects the latitude and longitude $(\lambda, \phi)$ on Jupiter to the pixels $(x, y)$ of the detector. The weighting maps $w_{zon,i}(\lambda, \phi)$ are determined in the same way from $\sigma_{v_{zon,i}}(x, y)$. Velocity maps $v(\lambda, \phi)$ are divided into elements of $\delta\lambda = 1°$ and $\delta\phi = 1°$. Then, each element is subdivided into a finer grid of $10 \times 10$ subelements of $(x, y)$. The corresponding velocities for all the subelements of $(x, y)$ are then determined by interpolation and averaged, providing the value of an element of $v_{zon,i}(\lambda, \phi)$. In parallel, we obtain a weight map (inverse of the noise) by summing the contribution of each added image. The resulting velocity map is then normalized by the sum of the weights $\sum_{i=1}^{N} w_{v_{zon,i}}(\lambda, \phi)$.

We repeated the procedure for the vertical and meridional components. The difference arises from the fact that projection factors are always positive with the longitude; therefore, any biases or noise are not canceled by the averaging over the longitudes, as in the case of the zonal maps. For instance, any variations of the zonal wind with time will not be canceled during the rotation of Jupiter and will result in spurious velocities in the meridional and vertical maps. In addition, for the meridional component, the region near the equator is affected by a strong noise because the projection factor becomes very small around the latitude of the subsolar and subterrestrial point. That is why we do not show the meridional component between $-6°$ and $+3°$ in latitude to avoid too-large values coming only from noise.

Last but not least, the vertical $V_{ver}$ and meridional $V_{mer}$ components are not independent since their projection factors only differ by $\tan\lambda$ (e.g., Gaulme et al. 2018). Actually, only the sum $V_{ver}\sin\lambda + V_{mer}\cos\lambda$ can be retrieved independently from the zonal contribution. The maps shown in Section 4.3 of the vertical (resp. meridional) components are the contribution to the Doppler measurement assuming the other component— meridional (resp. vertical)—to be null, which is certainly not the case. Therefore, the values of the vertical and meridional winds that we report need to be considered as upper limits. Moreover, we have seen that the correction to the Doppler measurement is very dependent on the intensity and velocity gradient and could only be corrected if we have an exact model of both the intensity and velocity field. Presently, the model used for the correction does not include any meridional contribution, except for a model of the GRS. These maps should then be interpreted with caution.

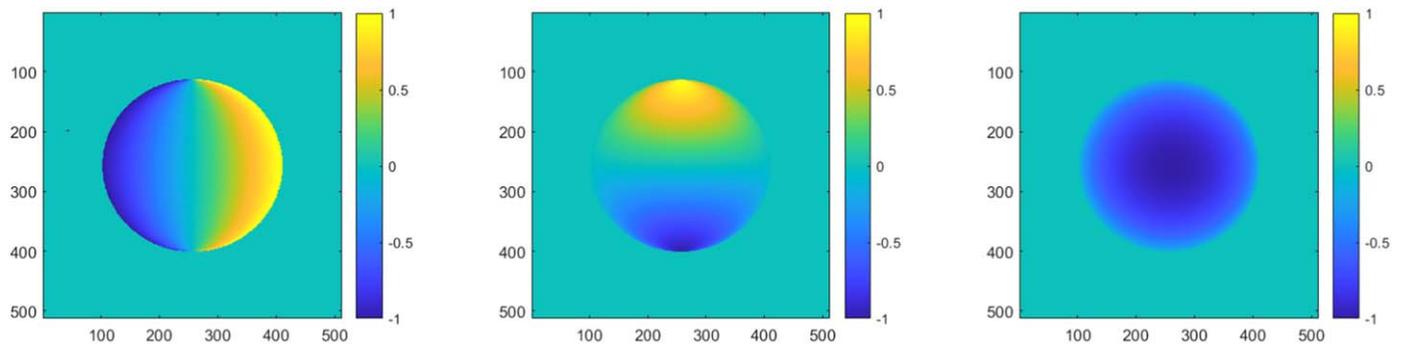

**Figure 12.** Projection factors of the Doppler velocity for the three components of the motion. The projection factors shown here are calculated for an oblate planet, with an inclination of 3°, Jupiter's typical inclination during the observations of 2018, and a phase of 4°, the maximum value during that period. The projection factors are calculated for the time of each image used in the treatment. Left: zonal component. Middle: meridional. Right: vertical motion.

## Appendix C
## Zonal Wind Profiles

Table 2 shows the values of the zonal wind profiles displayed in Figure 3.

**Table 2**
Zonal Velocity Profiles Obtained from the JIVE and HST OPAL Data that Are Presented in the Present Paper

| Planetocentric Latitude $\lambda$ (deg) | JIVE | | | HST/OPAL | |
|---|---|---|---|---|---|
| | Velocity $v_{\text{zon,JIVE}}$ (m s$^{-1}$) | Velocity Error (m s$^{-1}$) | Velocity Dispersion (m s$^{-1}$) | Velocity $v_{\text{zon,HST}}$ (m s$^{-1}$) | Velocity w/ Degraded Res. (m s$^{-1}$) |
| −60 | 3.57 | 3.36 | 54.29 | 2.07 | 1.33 |
| −59 | 1.66 | 3.19 | 51.57 | −0.17 | 2.35 |
| −58 | −0.52 | 3.05 | 49.49 | 0.95 | 3.54 |
| −57 | −4.08 | 2.92 | 47.36 | 2.98 | 4.88 |
| −56 | −7.90 | 2.80 | 44.79 | 4.31 | 6.38 |
| … | … | … | … | … | … |
| 59 | −28.84 | 7.68 | 134.22 | −0.34 | −1.46 |
| 60 | −51.62 | 11.95 | 222.13 | 3.41 | −1.09 |

(This table is available in its entirety in machine-readable form.)

## ORCID iDs

Francois-Xavier Schmider https://orcid.org/0000-0003-3914-3546
Patrick Gaulme https://orcid.org/0000-0001-8330-5464
Raúl Morales-Juberías https://orcid.org/0000-0001-7122-443X
Jason Jackiewicz https://orcid.org/0000-0001-9659-7486
Tristan Guillot https://orcid.org/0000-0002-7188-8428
Amy A. Simon https://orcid.org/0000-0003-4641-6186
Michael H. Wong https://orcid.org/0000-0003-2804-5086
Cristo Sanchez https://orcid.org/0009-0003-5699-5414
Riley DeColibus https://orcid.org/0000-0002-1647-2358
Sarah A. Kovac https://orcid.org/0000-0003-1714-5970
Sean Sellers https://orcid.org/0000-0001-5342-0701
Patrick Boumier https://orcid.org/0000-0002-0168-987X
Thierry Appourchaux https://orcid.org/0000-0002-1790-1951
Jean Pierre Rivet https://orcid.org/0000-0002-0289-5851
Steve Markham https://orcid.org/0009-0005-5613-3026
Saburo Howard https://orcid.org/0000-0003-4894-7271
Djamel Mekarnia https://orcid.org/0000-0001-5000-7292
Masahiro Ikoma https://orcid.org/0000-0002-5658-5971
Hidekazu Hanayama https://orcid.org/0000-0001-8221-6048
Bun'ei Sato https://orcid.org/0000-0001-7505-2487
Masanobu Kunitomo https://orcid.org/0000-0002-1932-3358
Hideyuki Izumiura https://orcid.org/0000-0002-8435-2569

## References

Atkinson, D. H., Ingersoll, A. P., & Seiff, A. 1997, Natur, 388, 649
Barrado-Izagirre, N., Rojas, J. F., Hueso, R., et al. 2013, A&A, 554, A74
Bolton, S. J., Levin, S. M., Guillot, T., et al. 2021, Sci, 374, 968
Buralli, B., Lai, O., Carbillet, M., et al. 2022, Proc. SPIE, 12185, 121858R
Cacciani, A., Dolci, M., Moretti, P. F., et al. 2001, A&A, 372, 317
Cacciani, A., & Fofi, M. 1978, SoPh, 59, 179
Choi, D. S., & Showman, A. P. 2011, Icar, 216, 597
Choi, D. S., Showman, A. P., Vasavada, A. R., & Simon-Miller, A. A. 2013, Icar, 223, 832
Civeit, T., Appourchaux, T., Lebreton, J.-P., et al. 2005, A&A, 431, 1157
Dahl, E. K., Chanover, N. J., Orton, G. S., et al. 2021, PSJ, 2, 16
Duer, K., Gavriel, N., Galanti, E., et al. 2021, GeoRL, 48, e2021GL095651
Fletcher, L. N., Oyafuso, F. A., Allison, M., et al. 2021, JGRE, 126, e2021JE006858
Fusco, T., Bacon, R., Kamann, S., et al. 2020, A&A, 635, A208
Galperin, B., Sukoriansky, S., & Huang, H.-P. 2001, PhFl, 13, 1545
Galperin, B., Young, R. M. B., Sukoriansky, S., et al. 2014, Icar, 229, 295
García-Melendo, E., & Sánchez-Lavega, A. 2001, Icar, 152, 316
Gaulme, P., Schmider, F. X., Gay, J., Guillot, T., & Jacob, C. 2011, A&A, 531, A104
Gaulme, P., Schmider, F. X., Gay, J., et al. 2008, A&A, 490, 859
Gaulme, P., Schmider, F.-X., & Gonçalves, I. 2018, A&A, 617, A41
Gaulme, P., Schmider, F.-X., Widemann, T., et al. 2019, A&A, 627, A82
Giles, R. S., Orton, G. S., Stephens, A. W., et al. 2019, GeoRL, 46, 1232
Gonçalves, I., Schmider, F.-X., Bresson, Y., et al. 2016, Proc. SPIE, 9908, 99083M
Gonçalves, I., Schmider, F. X., Gaulme, P., et al. 2019, Icar, 319, 795
Guillot, T., Miguel, Y., Militzer, B., et al. 2018, Natur, 555, 227
Guillot, T., Stevenson, D. J., Hubbard, W. B., & Saumon, D. 2004, in Jupiter. The Planet, Satellites and Magnetosphere, Vol. 1, ed. F. Bagenal, T. E. Dowling, & W. B. McKinnon (Cambridge: Cambridge Univ. Press), 35
Hsu, S., Crary, F. J., Parker, J., et al. 2019, AGUFM, P34C–02
Hueso, R., Sánchez-Lavega, A., Iñurrigarro, P., et al. 2017, GeoRL, 44, 4669
Ingersoll, A. P., Atreya, S., Bolton, S. J., et al. 2021, GeoRL, 48, e2021GL095756
Johnson, P. E., Morales-Juberías, R., Simon, A., et al. 2018, P&SS, 155, 2
Kaspi, Y., Galanti, E., Hubbard, W. B., et al. 2018, Natur, 555, 223
Lellouch, E., Paubert, G., Moreno, R., & Moullet, A. 2008, P&SS, 56, 1355
Limaye, S. S. 1986, Icar, 65, 335
Machado, P., Luz, D., Widemann, T., Lellouch, E., & Witasse, O. 2012, Icar, 221, 248
Machado, P., Silva, J. E., Brasil, F., et al. 2023, Univ, 9, 491
Machado, P., Widemann, T., Peralta, J., et al. 2017, Icar, 285, 8
Mosser, B., Maillard, J. P., & Mékarnia, D. 2000, Icar, 144, 104
Mosser, B., Mekarnia, D., Maillard, J. P., et al. 1993, A&A, 267, 604
Parisi, M., Kaspi, Y., Galanti, E., et al. 2021, Sci, 374, 964
Pepe, F., Cristiani, S., Rebolo, R., et al. 2021, A&A, 645, A96
Porco, C. C., West, R. A., McEwen, A., et al. 2003, Sci, 299, 1541
Read, P. L., Gierasch, P. J., Conrath, B. J., & Yamazaki, Y. H. 2005, AdSpR, 36, 2187
Read, P. L., Gierasch, P. J., Conrath, B. J., et al. 2006, QJRMS, 132, 1577
Salyk, C., Ingersoll, A. P., Lorre, J., Vasavada, A., & Del Genio, A. D. 2006, Icar, 185, 430
Savitzky, A., & Golay, M. J. E. 1964, AnaCh, 36, 1627
Schmider, F. X., Appourchaux, T., Gaulme, P., et al. 2013, in ASP Conf. Ser. 478, Fifty Years of Seismology of the Sun and Stars, ed. K. Jain et al. (San Francisco, CA: ASP), 119
Schmider, F. X., Fossat, E., & Mosser, B. 1991, A&A, 248, 281
Schmider, F. X., Gay, J., Gaulme, P., et al. 2007, A&A, 474, 1073
Seiff, A., Blanchard, R. C., Knight, T. C. D., et al. 1997, Natur, 388, 650
Shaw, C. L., Gulledge, D. J., Swindle, R., Jefferies, S. M., & Murphy, N. 2022, FrASS, 9, 768452
Simon, A. A., Hueso, R., Iñurrigarro, P., et al. 2018, AJ, 156, 79
Simon, A. A., Wong, M. H., & Orton, G. S. 2015, ApJ, 812, 55
Soulat, L., Schmider, F. X., Robbe-Dubois, S., et al. 2017, Proc. SPIE, 10564, 105641V
Sugiyama, K., Nakajima, K., Odaka, M., Kuramoto, K., & Hayashi, Y. Y. 2014, Icar, 229, 71
Tollefson, J., Wong, M. H., de Pater, I., et al. 2017, Icar, 296, 163
Underwood, T. A., Voelz, D., Schmider, F.-X., et al. 2017, Proc. SPIE, 10401, 104010Y
Vasavada, A. R., Ingersoll, A. P., Banfield, D., et al. 1998, Icar, 135, 265
Vorontsov, S. V., Zharkov, V. N., & Lubimov, V. M. 1976, Icar, 27, 109
Widemann, T., Lellouch, E., & Donati, J.-F. 2008, P&SS, 56, 1320
Wong, M. H., Marcus, P. S., Simon, A. A., et al. 2021, GeoRL, 48, e93982
Zirker, J. B. 1998, SoPh, 182, 1